\documentclass[
	aps, prl, reprint,
	a4paper,
	floats, floatfix,
	amsmath, amssymb, amsfonts,
	superscriptaddress, showpacs, footinbib
]{revtex4-1}

\newcommand{\rouge}[1]{\textcolor{red}{#1}}
\renewcommand{\rouge}[1]{}

\renewcommand{\section}[1]{\paragraph{#1. ---}\addcontentsline{toc}{section}{#1}}
\def\paper{Letter}

\usepackage{ifthen}
\newboolean{prd}
\setboolean{prd}{false}
\newboolean{arxiv}
\setboolean{arxiv}{true}

\newboolean{notprd}
\setboolean{notprd}{true}
\ifprd
\setboolean{notprd}{false}
\fi
\newboolean{notarxiv}
\setboolean{notarxiv}{true}
\ifarxiv
\setboolean{notarxiv}{false}
\fi

\usepackage{graphicx}
\usepackage[caption=false]{subfig}
\usepackage[usenames]{xcolor}

\ifnotprd
\xdefinecolor{mylinkcolor}{rgb}{0,0,0.5}
\usepackage[
	bookmarksnumbered, bookmarksopen, bookmarksopenlevel=2,
\ifnotarxiv
	breaklinks=true,
\fi
	colorlinks=true, filecolor=mylinkcolor, citecolor=mylinkcolor,
	linkcolor=mylinkcolor, urlcolor=mylinkcolor, menucolor=mylinkcolor,
]{hyperref}
\fi

\def\be{\begin{equation}}
\def\ee{\end{equation}}
\def\beq{\begin{eqnarray}}
\def\eeq{\end{eqnarray}}

\newcommand{\bea}{\begin{eqnarray}}
\newcommand{\eea}{\end{eqnarray}}
\newcommand{\ben}{\begin{enumerate}}
\newcommand{\een}{\end{enumerate}}
\newcommand{\bi}{\begin{itemize}}
\newcommand{\ei}{\end{itemize}}

\newcommand{\Ord}[2]{\mathcal O \left[#1\right]^{#2}}

\begin{document}

\title{New insights on the matter-gravity coupling paradigm}

\author{T\'erence Delsate}
\affiliation{CENTRA, Instituto Superior T\'ecnico, Avenida Rovisco Pais 1, 1049-001 Lisboa, Portugal, EU}
\affiliation{UMons, Universit\'e de Mons, Place du Parc 20, 7000 Mons, Belgium, EU}
\author{Jan Steinhoff}
\affiliation{CENTRA, Instituto Superior T\'ecnico, Avenida Rovisco Pais 1, 1049-001 Lisboa, Portugal, EU}

\date{\today}

\begin{abstract}
The coupling between matter and gravity in General Relativity is given by a proportionality relation between the stress tensor and the geometry. This is an oriented assumption driven by the fact that both the stress tensor and the Einstein tensor are divergenceless. However, General Relativity is in essence a nonlinear theory, so there is no obvious reason why the coupling to matter should be linear.
On another hand, modified theories of gravity usually affect the vacuum dynamics, yet keep the coupling to matter linear. In this {\paper} we address the implications of consistent nonlinear gravity/matter coupling. The Eddington inspired Born-Infeld theory recently introduced by Ba{\~n}ados and Ferreira provides an enlightening realization of such coupling modifications. 
We find that this theory coupled to a perfect fluid reduces to General Relativity coupled to a nonlinearly modified perfect fluid, leading to an ambiguity between modified coupling and modified equation of state. 
We discuss observational consequences of
this degeneracy and argue that such a completion of General Relativity is viable
from both an experimental and theoretical point of view through energy conditions, consistency, and singularity-avoidance perspectives. We use these results to discuss the impact of changing the coupling paradigm.
%

%

\end{abstract}

\pacs{04.50.-h, 04.20.Dw, 98.80.Bp, 04.80.Cc, 97.60.Jd, 04.25.Nx}

\maketitle

\section{Introduction}

Within Einstein's theory of General Relativity (GR), the existence of singularities,
e.g., at the big bang or within black holes, seems unavoidable
\cite{Hawking:Ellis:1973}. Worldlines of freely falling
observers come to an end within a finite proper time at such singularities. As a consequence,
it is not possible anymore to perform measurements within GR on such singularities, and an extending
theory is needed.
It is generally believed that quantum physics
will cure these problems. One usually admits that such fundamental theories 
would possess a classical effective
theory comprising some of its desired features.
In the past, candidate theories that soften singularity problems were studied~\cite{Feigenbaum:1998, *Wohlfarth:2003ss}, but all of these had to modify the vacuum field equations. However, recently Ba{\~n}ados and Ferreira discovered that it is possible to avoid the big bang singularity by effectively modifying the gravity/matter coupling (GraMaC) only~\cite{Banados:Ferreira:2010}. The specific theory proposed by Ba{\~n}ados and Ferreira~\cite{Banados:Ferreira:2010} was coined Eddington inspired Born-Infeld (EiBI) gravity. Interestingly, Born-Infeld electrodynamics~\cite{Born:1934gh} arises as a
low-energy effective theory of certain string theories~\cite{Fradkin:1985qd}, which is one of the original motivations to consider Born-Infeld inspired structures.

Our motivations originate somewhere else. Currently, the most stringent experimental tests of gravity probe spacetime in vacuum but not \emph{within} dense matter. As a consequence, the vacuum field equations are highly constrained, whereas the GraMaC is not. Therefore, there is a strong need to better understand implications of such modified coupling, both theoretically and experimentally, even if the latter is highly challenging. EiBI theory provides an optimal playground to understand implications of such a modified coupling. The reason is that EiBI coupled to a perfect fluid exactly corresponds to GR with a modified fluid equation of state (EOS). This allows an intuitive and explicit understanding of coupling modifications, avoiding technical difficulties. This leads to intuitive new interpretations of the recent results proposed
in~\cite{Banados:Ferreira:2010,Pani:Cardoso:Delsate:2011, Pani:Delsate:Cardoso:2012}.
Implications of modified coupling on energy conditions are illustrated within our framework. More important, we reveal the
singularity-avoiding mechanism discovered in~\cite{Banados:Ferreira:2010} and conjecture a similar effect in more general contexts.
We further stress that a bimetric-like action for EiBI theory provides a clean basis to independently modify the GraMaC and the vacuum field equations.

\section{EiBI as GR with apparent stress}
In this section we review the basics of EiBI theory and explicitly show the coupling modification in the case of a perfect fluid source.    
EiBI theory introduced in~\cite{Banados:Ferreira:2010} is based on a Palatini formulation
of the following action
\be
S = \frac{2}{\alpha \kappa}\int \! d^4x \! \left[ \sqrt{-\mbox{det}(g_{ab} + \kappa R_{ab})} - \lambda \sqrt{-g}\right] + S_M, \label{action}
\ee
where $g_{ab}$ is the metric, $g = \det(g_{ab})$, $R_{ab}$ is the Ricci tensor built from an independent connection $\Gamma_{ab}^c$, and $S_M$ is the matter action;
in~\cite{Banados:Ferreira:2010},  $\alpha=1$.
Note that most previous work on Born-Infeld-like gravity uses a metric formulation, see, e.g.,~\cite{Feigenbaum:1998, *Wohlfarth:2003ss, Deser:1998rj, *Gullu:2010em, Comelli:2004qr}.
Variation of \eqref{action} 
leads to~\cite{Banados:Ferreira:2010,Vollick:2003qp}
\bea
&&\lambda q_{ab} = g_{ab} + \kappa R_{ab}, \ \Gamma_{ab}^c = \frac{1}{2}q^{cd}\left( \partial_a q_{bd}+\partial_b q_{ad}-\partial_d q_{ab}\right),\nonumber\\
&&q^{ab} = \tau ( g^{ab} - \gamma \kappa T^{ab} ), \ \tau = \sqrt{\frac{g}{q}},\ T^{ab} = \frac{1}{\sqrt{-g}}\frac{\delta S_M}{\delta g_{ab}},
\label{field_eqs}
\eea
where $q_{ab}$ is an auxiliary metric, $q = \det(q_{ab})$, and $\gamma = \alpha / \lambda$. We have absorbed a factor of $\lambda$ in $q_{ab}$ compared to~\cite{Banados:Ferreira:2010}, both metrics are then identical if $T^{ab}=0$. 

Equations \eqref{field_eqs} may be written as
\be
q^{ac} g_{cb} = \lambda \delta^a{}_b - \kappa R^a{}_b, \ q^{ac} g_{cb} = \tau ( \delta^a{}_b - \gamma\kappa T^a{}_b ),
\ee
where $q^{ab},\ g^{ab}$ are the matrix inverse of $q_{ab}$, $g_{ab}$, respectively and $R^a{}_b = q^{ac}R_{cb}$, $T^a{}_b = T^{ac} g_{cb}$.
Combining both equations leads to
\be
R^a{}_b = \gamma\tau T^a{}_b + \frac{\lambda - \tau}{\kappa} \delta^a{}_b, \ R = \gamma\tau T + 4 \frac{\lambda - \tau}{\kappa} ,
\ee
where $R = R^a{}_a$ and $T = T^a{}_a$. The Einstein tensor for $q_{ab}$ then follows immediately,
\be
G^a{}_b[q_{cd}] = R^a{}_b - \frac{1}{2} \delta^a{}_b R = \gamma\mathcal{T}^a{}_b - \Lambda\delta^a{}_b ,
\label{einstein_q}
\ee
where we defined the apparent stress tensor $\mathcal{T}^a{}_b$ (the addition ``apparent''
will become clear in the next section),
\be
\mathcal{T}^a{}_b = \tau T^a{}_b + \mathcal{P} \delta^a{}_b , \
\gamma\kappa \mathcal{P} = \tau - 1 - \frac{1}{2} \tau \gamma\kappa T ,
\ee
and the cosmological constant $\Lambda = (\lambda - 1) / \kappa$.
$\mathcal{P}$ plays the role of an isotropic pressure addition.
Note that $\tau$ can be obtained from $T^a{}_b$ by
\bea
\tau &=& \left[ \det( \delta^a{}_b - \kappa \gamma T^a{}_b ) \right]^{-\frac{1}{2}} . \label{tau_def}
\eea
The Bianchi identities $G^{ab}{}_{|b}=0$ imply $\mathcal{T}^{ab}{}_{| b} = 0$,
where~${}_|$ denotes the covariant derivative based on $\Gamma_{ab}^c$.
The study of EiBI theory now reduces to a study of this potentially exotic source $\mathcal{T}^a{}_b$.

These expressions may still implicitly depend on $g_{ab}$ through $T^a{}_b$. 
However, in important special cases it is possible to completely decouple $g_{ab}$ from (\ref{einstein_q}).
For example, in the quite general case of a perfect fluid,
\be
T^a{}_b = \left( \rho + P  \right) u^a u_b + P \delta^a{}_b, \
u_a : u^a u^b g_{ab} = -1,
\ee
it is possible to reexpress $\mathcal{T}^a{}_b$ in the form of a perfect fluid, but in terms of $q_{ab}$, an apparent fluid velocity $v^a$, $v^av^bq_{ab}=-1$, and an apparent pressure $P_q$ and energy density $\rho_q$. In the following, we will refer to $g_{ab}$, $\rho$, and $P$ as the real metric, pressure, and energy density, while $q_{ab}$ will be referred to as the apparent metric.
After some algebra using the field equations \eqref{field_eqs}, one finds that $v_a v^b = u_a u^b$, while
$\tau$ is computed in a frame adapted to the fluid velocity,
\bea
&&\tau = \left[ (1 + \gamma\kappa\rho)(1-\gamma\kappa P)^3 \right]^{-\frac{1}{2}}.
\eea
It immediately follows that
\be
P_q = \tau P + \mathcal{P} , \
\rho_q + P_q = \tau (\rho + P),
\label{Pvirt}
\ee
and further
\be
\rho_q = \tau \rho - \mathcal{P} , \
\gamma\kappa\mathcal{P} = \tau - 1 - \frac{1}{2}\gamma \kappa \tau (3P-\rho) .
\ee

In the case of very dense matter, e.g., within neutron stars, the EOS can be
approximated to be barotropic, $\rho = \rho(P)$. Then (\ref{Pvirt})
encodes the apparent EOS. For example, dust $P=0$ leads to
\be
P_q=\frac{1}{3 \gamma  \kappa} \left[ 2 \sqrt{ 3 + ( 1 - \gamma  \kappa \rho_q )^2 } - 4 + \gamma\kappa\rho_q \right],
\ee
which smoothly interpolates between adiabatic index $2$ and $1$ for small and large, respectively, densities.
This provides intuitive interpretations for the existence of relativistic pressureless stars
and singularity avoidance for the collapse of dust in the Newtonian limit of EiBI theory~\cite{Pani:Cardoso:Delsate:2011, *Pani:Delsate:Cardoso:2012}.
(The vacuum energy EOS, $\rho = - P$, is invariant.)
Yet if a more realistic microphysics is desired, one has to find apparent counterparts for other thermodynamic variables as well.
Fortunately, a scalar quantity $n$ (e.g., baryon number density) with associated conserved current $j^a = n u^a$ transforms in an elegant way. Indeed, the conservation equation $\partial_a \sqrt{-g}j^a = 0$ can be rewritten as $\partial_a \sqrt{-q} n_q v^a = 0$, where $n_q = - n \tau / (v^b u_b)$. The term $- v^b u_b = \sqrt{\tau (1 + \gamma\kappa\rho)}$ is the time shift between $g_{ab}$ and $q_{ab}$, so the total factor $-\tau/(v^b u_b)$ is the ratio of the comoving-frame 3-volume measured with $g_{ab}$ or $q_{ab}$.

A minimally coupled scalar field with Lagrangian density $\mathcal L_m = [(\partial_a\phi)(\partial^a\phi)/2 - V]$ provides a specific microphysics for a perfect fluid. The resulting stress tensor~\cite{Madsen:1988, *Faraoni:2012hn} admits $P = \frac{\mathcal L_m}{2},\ \rho = \frac{\mathcal L_m}{2} +V$, where $\phi$ can be thought as a parameterizing function. One can then immediately compute the equations of motion in terms of $P_q$ and $\rho_q$.
It should be stressed that the scalar field pressure and density implicitly depend on $g_{ab}$, but it is possible to completely eliminate it although leading to very cumbersome expressions.

\section{Observational consequences}

Provided one relies on gravitational/geometrical imprints measured \emph{outside} matter, one needs to make assumptions on the  GraMaC to draw conclusions on the source. Remember that EiBI theory agrees with GR in vacuum and only modifies the GraMaC. If the GR coupling is falsely assumed, then the apparent quantities, e.g., the matter EOS, are interpreted as real.
This is the reason why we qualify these quantities as ``apparent.'' Thus by observations
of mass-radius relations or quasinormal mode frequencies of neutron stars, gravitational waves in general,
and so on, one may only gather information about the \emph{apparent} EOS.
For similar reasons, we identify $\gamma=8\pi G$ here, with the Newton constant $G$.

On the other hand, the \emph{real} EOS of very dense matter is not fully under control yet~\cite{Baldo:2011gz}.
Additionally, probing spacetime \emph{within} such highly interacting matter
seems to be extremely difficult.
However, an important experimental signature of EiBI theory might be given by the
case when gravitational waves favor a specific EOS, while observations
of electromagnetic fields or neutrinos favor a different EOS (multi-messenger astronomy~\cite{Marka:2011zz, *Pradier:2010mr}).
Remember that the latter couple to the real metric within the action. Finally,
it seems inevitable to constrain the real EOS by experiments on earth (based on electromagnetic interactions)
and/or by reaching consensus in theoretical predictions.

It is illustrative to constrain EiBI theory using the sun~\cite{Casanellas:2011kf}, though better constraints
are provided in~\cite{Avelino:2012ge}. Yet EiBI theory possesses excellent experimental viability.
Notice that the dynamics of compact binaries is almost identical to the GR
case, as apparent relative position and masses exactly equate the
real ones. Merely subleading EOS-dependent effects arise, which appear in
post-Newtonian frameworks as, e.g., tidal interactions~\cite{Damour:Soffel:Xu:1991, *Baiotti:etal:2010}.

\section{Energy conditions}

The most important implications of matter properties in GR are
encoded in various energy conditions. These have immediate consequences on, e.g.,
singularity theorems~\cite{Hawking:Ellis:1973}, entropy bounds~\cite{Bousso:2002}, and instabilities or other pathologies
\cite{Buniy:Hsu:2005, *Visser:Carlos:1999}.
The energy conditions for perfect fluids are expressed simply by the following requirements on pressure and density: 
\bea
&&\mbox{Null Energy Condition: }\rho+P\ge0,\nonumber\\
&&\mbox{Weak Energy Condition: } \rho+P\ge0,\ \rho\ge 0,\nonumber\\
&&\mbox{Strong Energy Condition: }\rho + P \ge 0,\ \rho+3P\ge0,\nonumber\\
&&\mbox{Dominant Energy Condition: } \rho\ge|P|,\nonumber\\
&&\mbox{Causal Energy Condition: } |\rho|\ge|P|.\nonumber
\eea

Since $\mathcal{T}^a{}_b$ is again a perfect fluid, we can check if the apparent EOS still fulfills a specific energy condition when the real EOS does. In the following, we refer to the energy condition expressed in terms of the apparent quantities as ``apparent energy conditions'' and, correspondingly, those expressed with the real quantities as ``real energy conditions.''

\begin{figure}
 \includegraphics[scale=.2553]{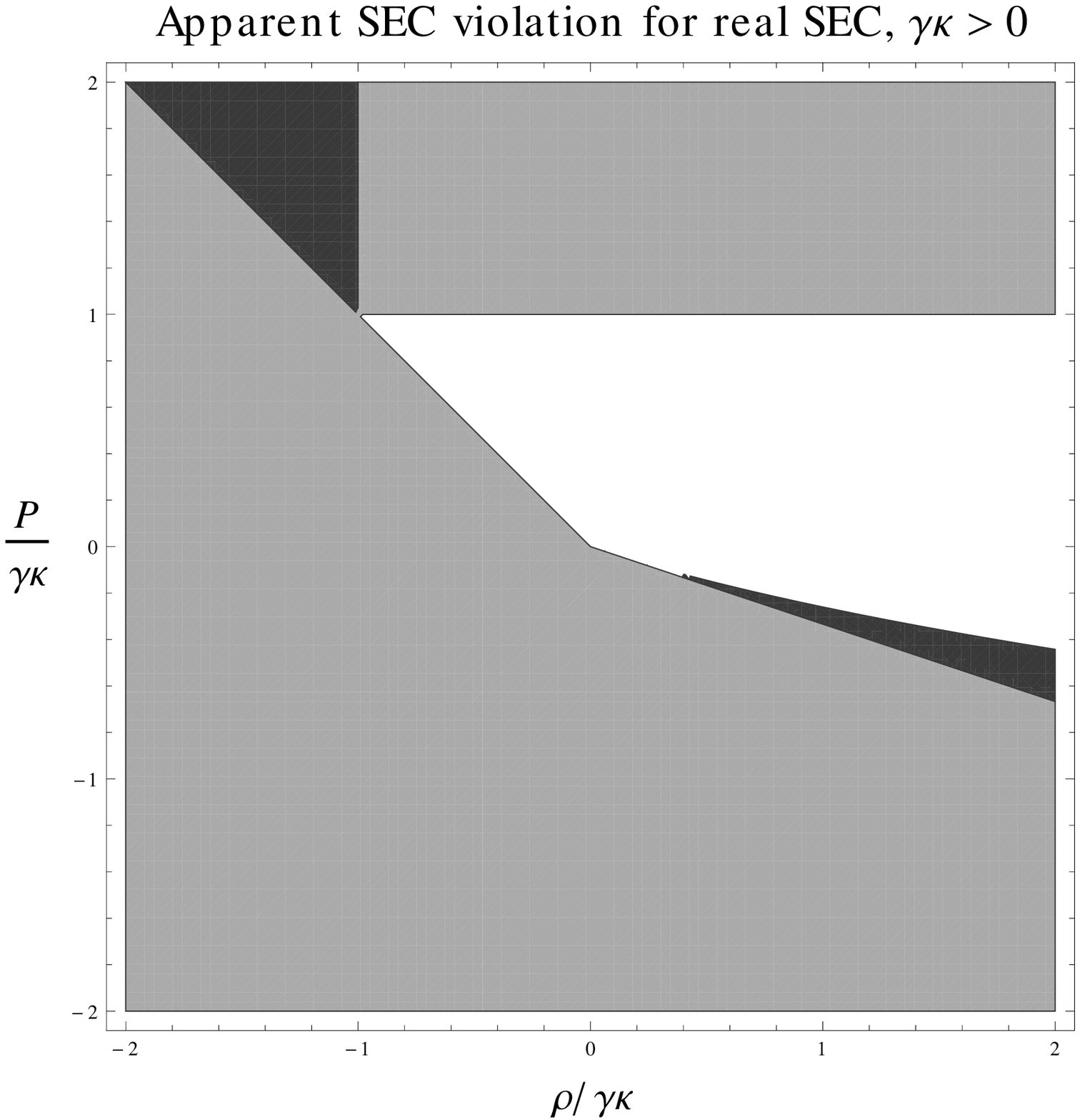} \hspace{-.3cm}\includegraphics[scale=.245]{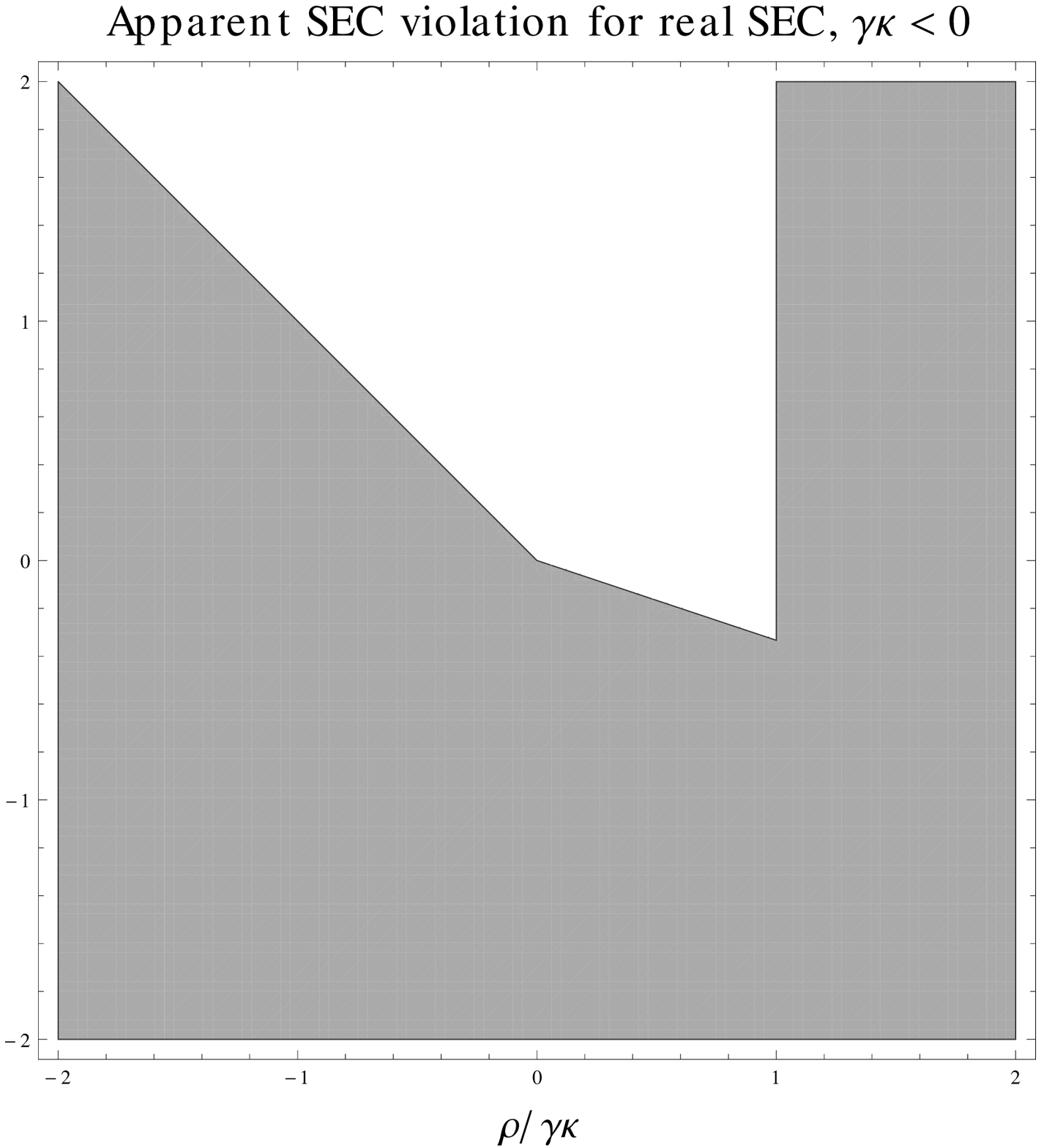}
\caption{{\bf Left:} Region excluded by real SEC or imaginary $\tau$ (light gray) and where real SEC leads to apparent SEC violation (dark gray) for $\kappa>0$ in the $\rho-P$ plane.  {\bf Right:} Same plot for $\kappa<0$.}
\label{fig:sec}
\end{figure}

The apparent Null Energy Condition (NEC) under the assumption of real NEC for perfect fluids follows from \eqref{Pvirt} and $\tau \geq 0$. Apart from NEC, energy conditions do not generically hold. Indeed, assuming real Strong Energy Condition (SEC) does not lead to apparent SEC. We illustrate this for $\kappa>0$ and $\kappa<0$ in Fig.\ \ref{fig:sec}.
As can be seen, for $\kappa>0$ there is indeed a dark gray region where assuming real SEC, we can find apparent SEC violation.
The SEC ensures by virtue of the Raychaudhuri equation that gravity is attractive (for $\Lambda = 0$) and
is therefore important for certain singularity theorems~\cite{Hawking:Ellis:1973}.
Yet, singularity avoidance occurs for both signs of $\kappa$, so it is not due to apparent SEC violation.

One can easily check that Weak Energy Condition (WEC) is transported to the apparent sector for positive values of $\kappa$ but can be violated for $\kappa<0$. Dominant and Causal Energy Conditions (DEC/CEC) can be violated for both signs of $\kappa$. This shows the possibility of apparent causality violations via superluminal energy flows inside matter~\cite{Hawking:Ellis:1973, Bousso:2002}. (But superluminal speed of sound is not excluded even by DEC or CEC.)

\section{Cosmological case revised}

An instructive situation considered in~\cite{Banados:Ferreira:2010} is the cosmological case. 
Depending on the sign of $\kappa$, the early time, radiation dominated universe either freezes or experiences a bounce, while the (real) density and pressure remain finite.
We have rederived these results.
Interestingly, although the real quantities are regular at all time, we find that the apparent sector \emph{does have} singularities, as predicted by singularity theorems.

\begin{figure}
 \includegraphics[scale=.3575]{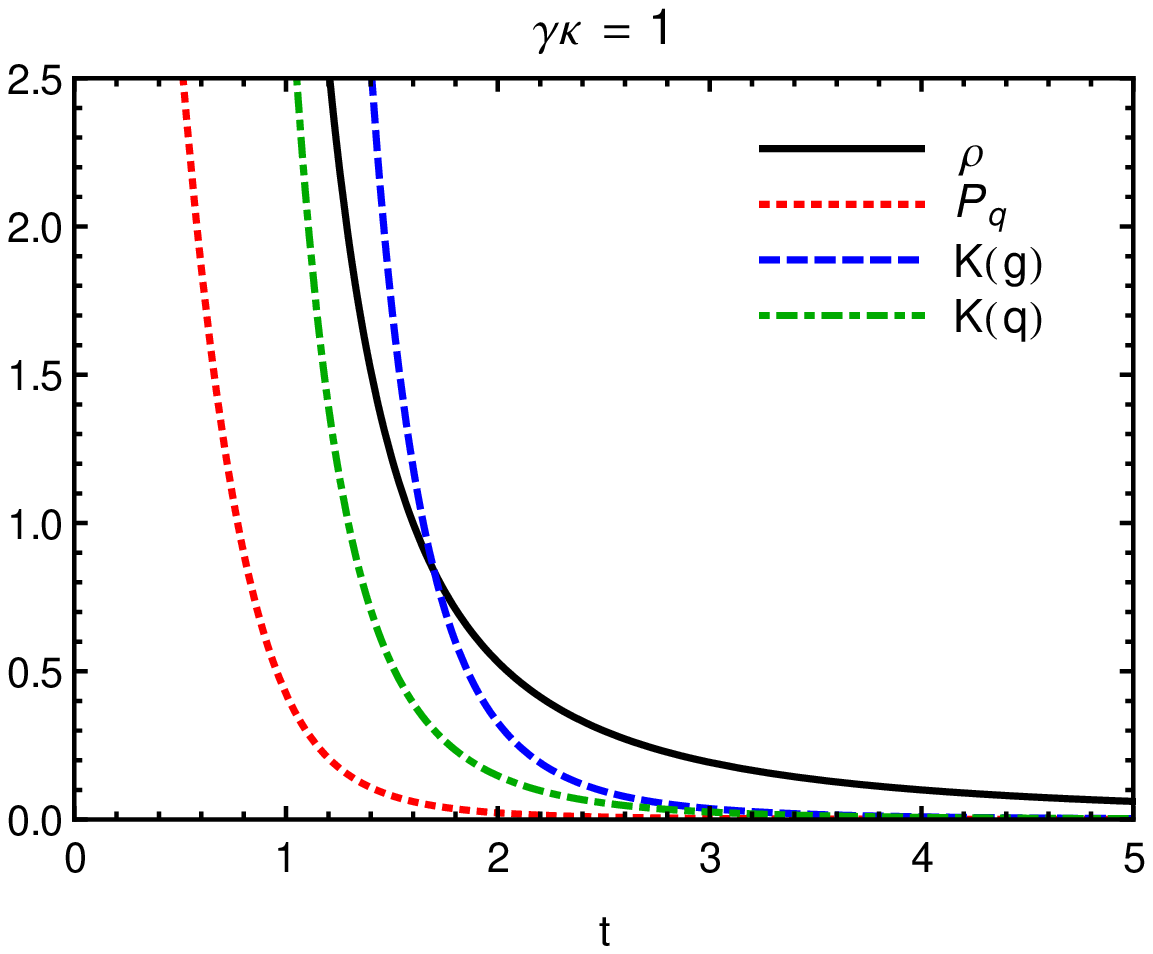} \hspace{-.3cm}\includegraphics[scale=.3575]{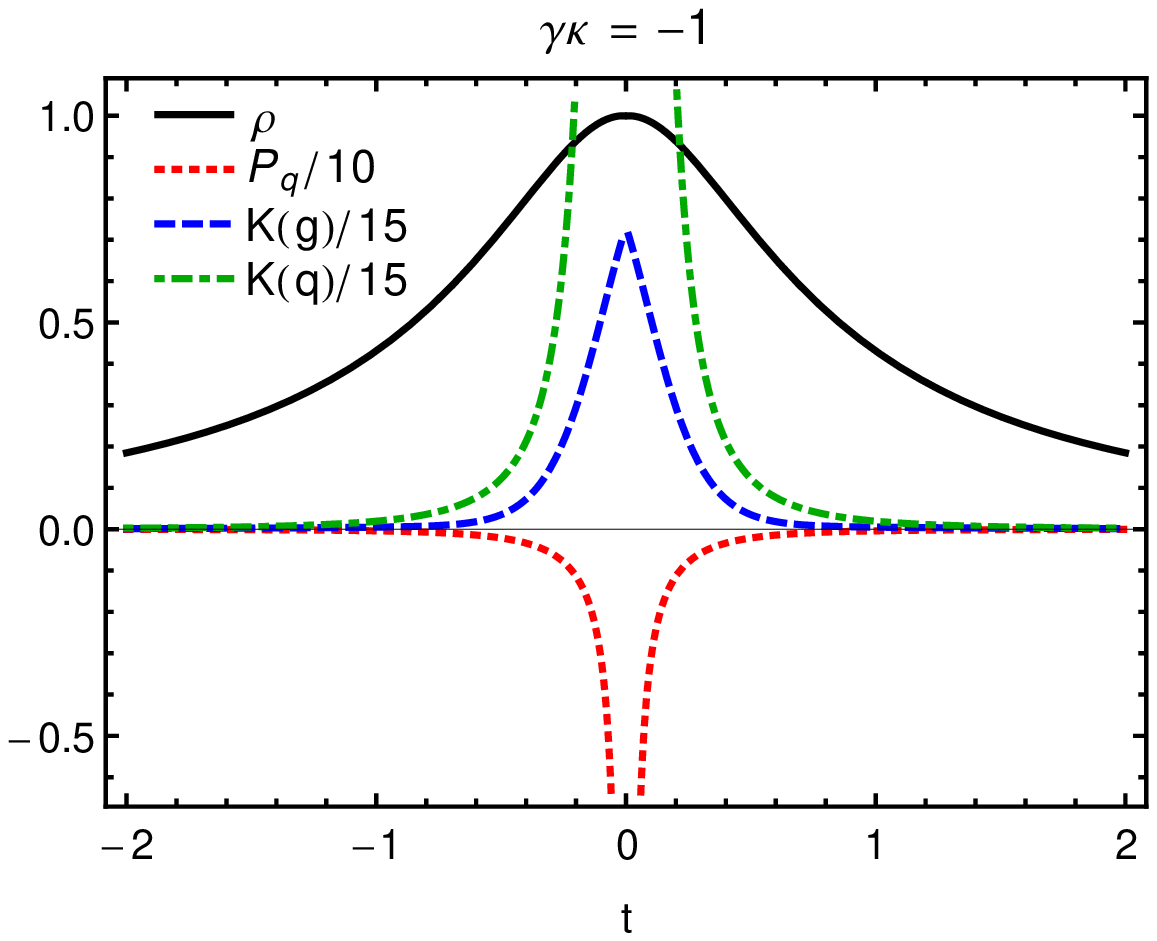}
\caption{Various quantities as a function of time $t$ for $\kappa>0$ (left) and $\kappa<0$ (right) for a dust universe, $P=0$.}
\label{fig:apparent}
\end{figure}

However, the case of a dust universe $P=0$ reveals that singularity avoidance is
not a generic feature of the theory. As illustrated in Fig.\ \ref{fig:apparent},
for $\kappa>0$, the Kretschmann scalar $K$ signals a singularity
in \emph{both} real and apparent sectors, while for $\kappa<0$ a bounce with regular real
sector occurs (analogous to the radiation universe~\cite{Banados:Ferreira:2010}).
This is opposite of what one might expect from the Newtonian collapse~\cite{Pani:Cardoso:Delsate:2011, *Pani:Delsate:Cardoso:2012}.
An intriguing explanation is given in the following.


\section{Singularity avoidance}

Interestingly, $\tau$ may become infinite for finite values of
$\rho$ and $P$. In the present section we argue that this triggers singularity avoidance.

The first case where  $\tau \rightarrow \infty$ is given by $\gamma \kappa P \rightarrow 1$.
Then both $P_q / \tau$ and $\rho_q / \tau$ approach a finite constant,
so $P_q$ and $\rho_q$ become infinite. This means that infinite
$P_q$ and $\rho_q$ correspond to finite $\rho$ and $P$. More precisely, $P=1/\gamma \kappa$ is the maximal
real pressure (corresponding to a singularity of $\rho_q$). Further, the apparent EOS
approaches $P_q \approx \rho_q$, independent of the real EOS, $\kappa$, or other thermodynamic variables.
A neutron star EOS is therefore considerably softened, which qualitatively
explains the modified mass-radius relation in~\cite{Pani:Cardoso:Delsate:2011, *Pani:Delsate:Cardoso:2012}
for $\kappa>0$.
Note that the time shift between apparent and real cosmological times $- v^b u_b $ approaches $\infty$.
Thus, the apparent singularity does not form within finite real time.

Similarly, the second case where $\tau \rightarrow \infty$ is given by
$\gamma \kappa \rho \rightarrow - 1$.
Again $P_q / \tau$ and $\rho_q / \tau$ approach a constant, but now
the apparent EOS approaches $P_q \approx - \rho_q / 3$ and
the maximal ($\kappa<0$) real energy density is $\rho = - 1/\gamma \kappa$.
Note that now the time shift goes to zero.

Whenever one of these two cases applies, gravity is not able to produce singular $\rho$ and $P$. But it is easy to give examples where both cases are avoided, e.g., the dust universe with $\kappa>0$ explored in the last section. However, most realistic EOS will
imply that either $\gamma \kappa P \rightarrow 1$ (for $\kappa > 0$) or $\gamma \kappa \rho \rightarrow - 1$
(for $\kappa < 0$) in the high density regime.
Thus, even when the apparent EOS fulfills the SEC and singularity theorems predict
a singularity in the apparent theory, $\rho$ and $P$ remain finite
in realistic situations.
This suggests that a regularization analogous to the cosmological case might occur for black hole formation, at least within matter. But further analysis is needed to understand under which circumstances worldlines of free falling observer will enter the matter and not end within finite real proper time, thus avoiding singularities. Notice that the diverging time shift for $\kappa>0$ indicates that not even the singularity in the apparent sector is encountered by an observer within finite real proper time.


\section{Ghosts, instabilities, and truncations}

Recently, unitarity analysis was performed for curvature-truncated Born-Infeld gravity actions~\cite{Deser:1998rj, *Gullu:2010em}. The result is a generic appearance of ghosts in the theory around a constant curvature background. These results were found within a metric approach, whereas here we follow a Palatini approach. Roughly speaking, ghosts are due to a ``wrong sign'' in the kinetic part of the action, which in particular is associated with instabilities of the theory. The gravitational part of (\ref{action}) can be \emph{equivalently} written as a linear bimetric-like action,
\be
S_G =\frac{1}{\gamma}\int \! d^4x \sqrt{-q} \left[ R[q]  - 2 \frac{\lambda}{\kappa} + \frac{1}{\kappa} ( q^{ab}g_{ab} - 2 \tau ) \right] \! ,
\label{equiv_action}
\ee
which is basically the action in~\cite{Banados:2008fi, *Banados:2008fj} without dynamics for $g_{ab}$.
This action is manifestly free of ghosts. 

NEC violation may lead to other types of instabilities or pathologies
\cite{Buniy:Hsu:2005, *Visser:Carlos:1999}. Assuming real NEC in the full nontruncated theory preserves apparent NEC, avoiding this type of instabilities. Instead, at the truncated level, it is very easy to violate NEC even at second order in truncation,
\be
\rho_q+P_q\approx (\rho+P) \left[1+\frac{\gamma\kappa}{2}\left( 3 P -\rho \right)\right] + \Ord{\rho,P}{2}.
\ee
A straightforward example of NEC violation within the truncated theory is $P=-\rho+\epsilon^2$, $\epsilon\in\mathbb R$, $\kappa>0$, for any $\rho>(2+3\gamma\kappa \epsilon^2)/(4\gamma\kappa)$.
One can equivalently truncate the curvature sector in the original action (\ref{action}).
This makes apparent that truncating a Born-Infeld gravity theory can lead to problems, whereas the nontruncated version is well behaved.

Truncations also ruin singularity avoidance, e.g., as for a post-Newtonian
approximation of the apparent source,
\be
P_q = P + \frac{\gamma\kappa}{8} \rho^2 + \dots, \
\rho_q = \rho - \frac{5\gamma\kappa}{8} \rho^2 + \dots .
\ee
Still this recovers the repulsive effect of EiBI theory for $\kappa > 0$ found
in~\cite{Banados:Ferreira:2010, Pani:Cardoso:Delsate:2011, *Pani:Delsate:Cardoso:2012}.
Indeed, for $\kappa > 0$ the apparent theory possesses both an increased
pressure and a decreased density compared to the real theory.
[But note that the $\kappa$-dependent term in (7) of~\cite{Banados:Ferreira:2010} is coming from the relation between $q_{ab}$ and $g_{ab}$.]

\section{Conclusions and outlook}

EiBI theory provides a prototype of GraMaC modification at high densities in a way that avoids the big bang
singularity~\cite{Banados:Ferreira:2010}.
EiBI theory reproduces usual GR in vacuum and to lowest order in $\kappa$. 
Let us stress further remarkable features of EiBI theory
discovered in the present {\paper} and how they translate to more general modifications of GraMaC.


We found that in the case of a perfect fluid, EiBI theory \emph{exactly} reduces
to GR, but in terms of an apparent metric coupled to a fluid with
an apparently modified EOS. The notion of apparent quantities generalizes to more generic GraMaC, although real fluids will not always map to apparent fluids. 
On the one hand, this presents a challenge for
experimental constraints on the theory (and concurrently on the real nuclear EOS from neutron star observations), since it is very difficult to probe spacetime within matter.
On the other hand, this demonstrates the excellent current and future experimental viability of modified GraMaC theories.
Notice that EiBI theory is viable also from a theoretical point of view, as it
appears to be free of instabilities, ghosts, or other inconsistencies. Such theories can also make regular matter appear exotic as illustrated by the energy conditions.

The similarity to GR allows for a transfer of the vast amount of knowledge
on GR to EiBI theory in the future, e.g., usage of existing codes for full
numerical simulations, exploration of the Oppenheimer-Snyder solution~\cite{oppenheimer:snyder:1939} for the
collapse of dust (where now $P_q = 0$), or study of naked apparent singularities
as hypothetical astrophysical objects~\cite{Joshi:2012mk}. This should also be the case in more general situations since essentially the description of the source needs to be reimplemented.

We revealed that singularity avoidance in EiBI theory occurs when a pole in
$\tau$, Eq.\ (\ref{tau_def}), is approached and illustrated this mechanism by \emph{singular and nonsingular}
examples in the cosmological case. Though in the nonsingular case the apparent
curvature becomes singular, this is completely harmless for measurable (real) quantities.
It is most interesting to study explicitly the geodesics or the validity of entropy bounds
during a collapse to a black hole in the future.
It can be expected in some particular cases EiBI that allows physical predictions where GR fails.


An intriguing new interpretation of the gravitational interaction derives from \eqref{equiv_action},
in particular when viewed as a rather unusual effective action.
This action consists of the ``bare'' Einstein-Hilbert part for the dynamical metric $q_{ab}$, which induces
a dynamics to the measurable (matter-coupled) metric $g_{ab}$ through a constraint-type relation.
The ``cutoff'' $(\gamma\kappa)^{-1}$ enters \eqref{equiv_action} as a coupling constant.
Instead, in bimetric theories, usually both metrics are truly dynamic. This makes the action \eqref{equiv_action} interesting as a basis for independent modification of GraMaC and vacuum field equations, opening up a new class of modified theories of gravity.

\acknowledgments
We gratefully acknowledge fruitful discussions with P. Pani, V. Cardoso, and V. Vitagliano.
This work was supported by DFG (Germany) through project STE 2017/1-1,
FCT (Portugal) through projects PTDC/CTEAST/098034/2008 and  PTDC/FIS/098032/2008 and CERN through project 
CERN/FP/123593/2011.

\ifnotprd
\bibliographystyle{utphys}
\fi

\ifarxiv
\providecommand{\href}[2]{#2}\begingroup\raggedright\endgroup

\else
\bibliography{NS_Eddi_tidal}
\fi

\end{document}